\documentclass[twocolumn,showpreprint,preprintnumbers,amsmath,amssymb]{revtex4}
\usepackage{amssymb}
\usepackage{txfonts}

\usepackage{txfonts,amssymb,textcomp,graphicx,epsfig}
\usepackage{amsmath}

\begin{document}

\preprint{}

\title{Strain-induced change in
local structure and its effect on the ferromagnetic properties of
La$_{0.5}$Sr$_{0.5}$CoO$_3$ thin films}

\author{C. K. Xie}
\author{J. I. Budnick}
\author{W. A. Hines}
\author{B. O. Wells}%
\affiliation{Department of Physics, University of Connecticut, Storrs, CT 06269}%

\author{J. C. Woicik}%
\affiliation{National Institute of Standards and Technology,
Gaithersburg, Maryland 20899}%

\date{\today}

\begin{abstract}
We have used high-resolution Extended X-ray Absorption
Fine-Structure and diffraction techniques to measure the local
structure of strained La$_{0.5}$Sr$_{0.5}$CoO$_3$ films under
compression and tension. The lattice mismatch strain in these
compounds affects both the bond lengths and the bond angles, though
the larger effect on the bandwidth is due to the bond length
changes. The popular double exchange model for ferromagnetism in
these compounds provides a correct qualitative description of the
changes in Curie temperature $T_C$, but quantitatively
underestimates the changes. A microscopic model for ferromagnetism
that provides a much stronger dependence on the structural
distortions is needed.
\end{abstract}

\maketitle


Epitaxial strain in thin films is often used to modify a material's
physical properties and improve device performance. For example,
biaxial strain can introduce bond-length and bond-angle distortions
in semiconductor alloys, \cite{Woicik2,Woicik3,Woicik4,Maiti}which
greatly affect their performance in real applications. Room
temperature ferroelectricity has been induced by lattice strain in
SrTiO$_3$thin films, a material that is not ferroelectric in the
bulk~\cite{jh04,woicik}. Enhanced magnetoresistance has been
achieved in La$_{0.8}$Ba$_{0.2}$MnO$_3$ thin films at room
temperature~\cite{zhang}, which makes it a potential candidate for
magnetic devices and sensors. The modification of physical
properties using strain is also an important tool for understanding
the physics of correlated electron materials. One longstanding
question in the field is the origin of ferromagnetism in several
poorly conducting transition-metal oxides. The most popular model
has been Zener's double exchange mechanism (DE)~\cite{zener}. In
this paper we report results comparing the Curie Temperature with
detailed structural measurements in strained films of
La$_{0.5}$Sr$_{0.5}$CoO$_3$~(LSCO). While the predictions appear
qualitatively correct, they do not quantitatively predict the
correct dependence on lattice parameter and, therefore, bandwidth.
Thus, either another mechanism or a modification to DE is needed.

The perovskite, transition-metal oxide that has been most studied as
a function of strain is the colossal magnetoresistive
manganites~\cite{cm00}. The strain has been induced in several ways
in manganites including films. The analysis of these experiments has
examined how strain has mediated the ferromagnetic coupling via
modification of the bandwidth $W$~\cite{zener,pw55}. In the
tight-binding model, the bandwidth W depends on the overlap
integrals between the Mn 3d and O 2p orbitals such that $W \propto
d^{-3.5}\cos\omega$~\cite{wa89,mm95}, where $d$ is Mn-O bond length,
$\omega=(180^{\circ}-\phi)/2$ is the tilt angle, and $\phi$ is
Mn-O-Mn bond buckling. A decrease in the lanthanide ion radius by
chemical substitution leads to a reduction of $T_C$ that has been
attributed to an increase of the Mn-O-Mn bond angle with little
change in the Mn-O bond length~\cite{Hwang, Radaelli}. The opposite
case is that compressive hydrostatic pressure increases $T_C$ due to
a reduction in Mn-O bond length with little decrease in Mn-O-Mn bond
angle~\cite{dp04}. However, for film studies no full consensus has
been reached. Yuan~\cite{Yuan} proposed that tensile strain
primarily increases the Mn-O-Mn bond angle to explain the
enhancement of $T_C$. Miniotas et al.~\cite{am01} reported that the
Mn-O bond length remains fixed, while the Mn-O-Mn bond angle
accommodates the strain. Similar behavior was assumed in other
work~\cite{hl98}. However, an x-ray absorption spectroscopy (XAS)
study indicated an energy shift at Mn K-edge in strained
La$_{0.7}$Sr$_{0.3}$MnO$_3$ thin films, implying a variation of Mn-O
bond length~\cite{sn04}. The system of La$_{1-x}$Sr$_x$CoO$_3$ is of
great interest in its own rite, with a wide variety of
applications~\cite{ss98,mr98,sh01}. Recent studies have shown that
strain can directly alter the ferromagnetic exchange coupling
energy~\cite{xie07}. The details of just how this happens require
more exact measurements of the local structure as reported here
using extended x-ray absorption fine structure (EXAFS).


LSCO films with thickness of 22 nm were epitaxially grown by a
pulsed laser deposition technique on LaAlO$_3$ (LAO) and SrTiO$_3$
(STO) substrates. X-ray diffraction (XRD) confirmed the epitaxy of
the films and indicated the absence of impurity phases or grains
with other orientations. We determined both the in-plane and
out-of-plane lattice strains by $\varepsilonup$=(a$_f$-a$_b$)/a$_b$,
where the a$_f$ and a$_b$ ($\simeq 3.834\AA$) are the lattice
constants for LSCO film and bulk, respectively, as measured by XRD.
The field-cooled DC magnetic properties were measured using a
superconducting quantum interference device magnetometer. High
resolution EXAFS experiments were conducted on the Co K edge at room
temperature using the National Institute of Standards and Technology
beamline X23A2 at the National Synchrotron Light Source, Brookhaven
National Laboratory. The absorption data were collected in two
orientations, with the polarization vector
$\overrightarrow{\varepsilon}$ of the synchrotron radiation aligned
either parallel or perpendicular to the sample surface. The x-ray
absorption spectrum from finely ground LSCO powder measured in
transmission was used as the EXAFS phase and amplitude standard to
experimentally determine the Co-O bond lengths within the films.

\begin{figure}
\centering
\epsfig{file=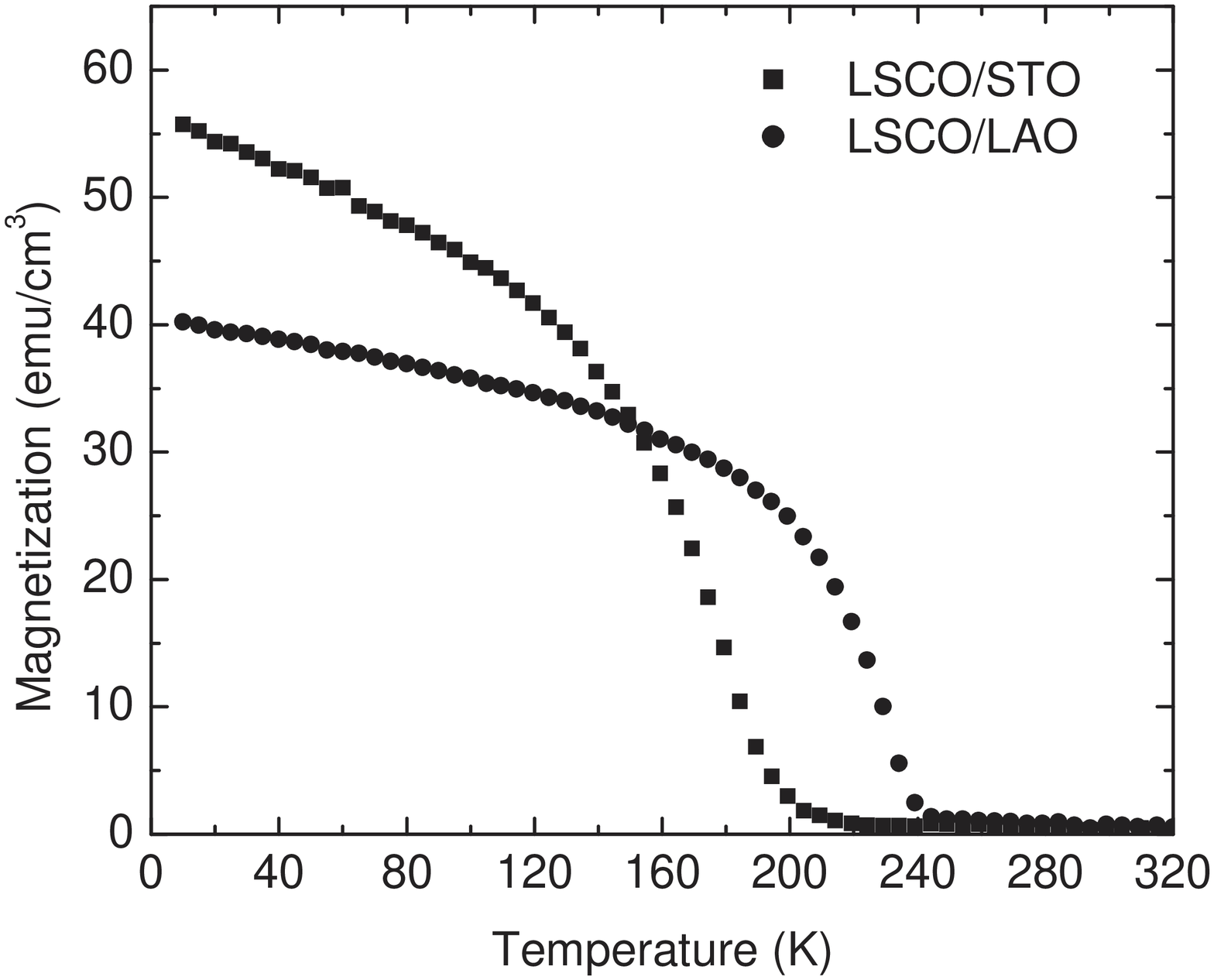,width=7cm,clip=}
\caption{\label{fig:fig1.eps} Temperature dependence of field-cooled
magnetization with applied field H= 100 Oe for LSCO films on STO and
LAO substrates.}
\end{figure}

\begin{figure}
\centering
\epsfig{file=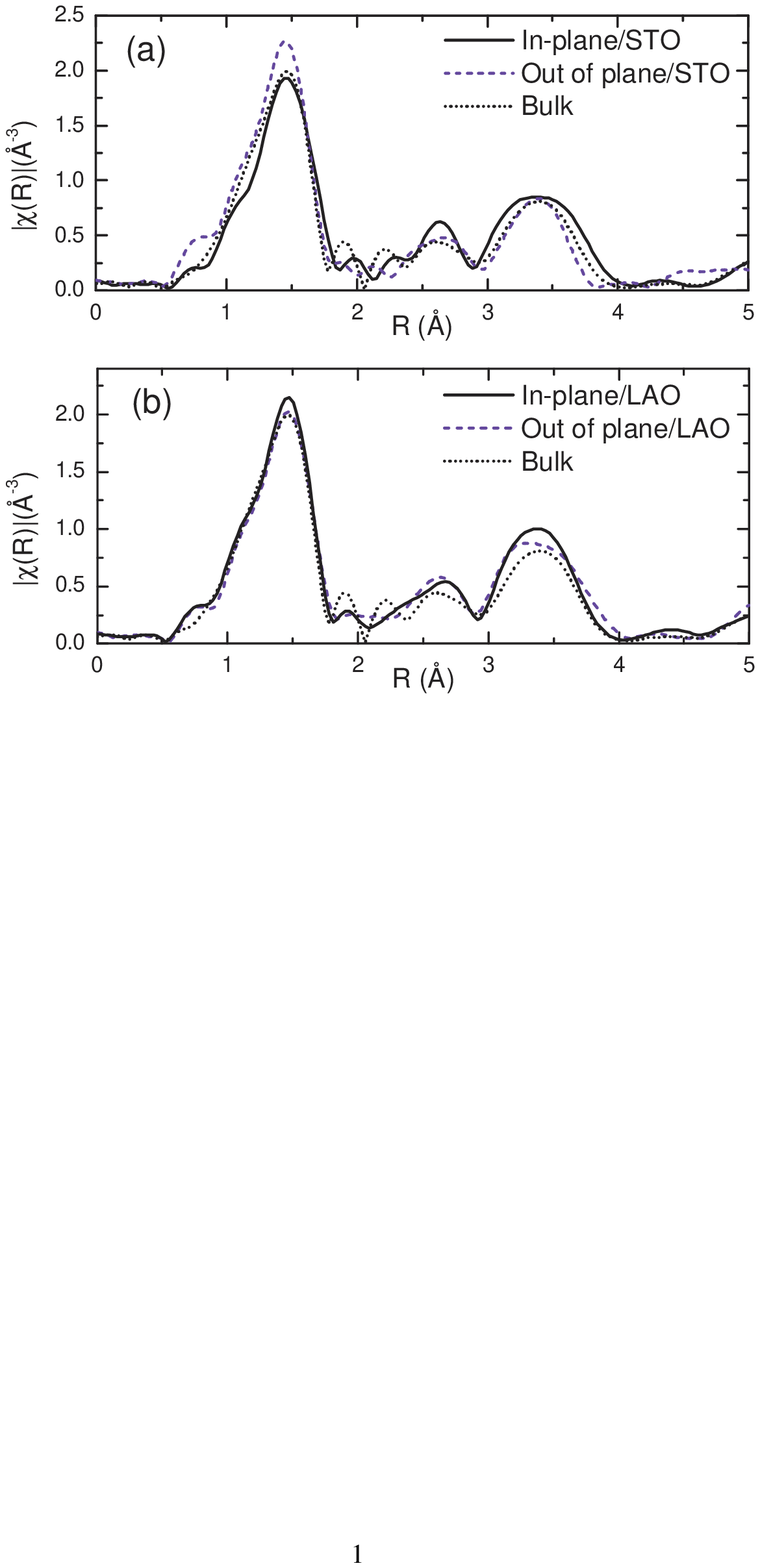,width=7cm,clip=}

\caption{\label{fig:fig2.eps} Magnitude of the Fourier transforms
(FT) of the k$^2$-weighted Co K-edge EXAFS with the polarization
vector of the synchrotron radiation aligned in-plane and
out-of-plane for LSCO thin films grown on (a) STO, and (b) LAO
substrates, respectively. For comparison, data of LSCO powder as a
standard are also shown.}
\end{figure}

Figure 1 shows the field-cooled magnetization curves M(T) for LSCO
thin films grown on STO and LAO substrates, respectively. The film
on STO has in-plane lattice constant 3.891 $\pm$ 0.002~\AA~ and
out-of-plane lattice constant 3.778 $\pm$ 0.001~\AA~ to give
in-plane tensile strain of +1.49\%, out-of-plane compressive strain
of -1.46\%, and $T_C$ of 198 K. The film on LAO has in-plane lattice
constant 3.808 $\pm$ 0.004~\AA~ and out-of-plane lattice constant
3.869 $\pm$ 0.002~\AA~ to give in-plane compressive strain of
-0.68\%, out-of-plane tensile strain of +0.91\%, and $T_C$ of 241 K.
Errors in lattice constants were estimated based on the standard
deviation in the peak width from fits to the XRD peaks. These $T_C$
values should be compared to the expected, unstrained value for
$T_C$ of 229K determined in our previous work, which is itself
reduced from the bulk value of 250 K due to the finite size
effect\cite{xie07, sm96}. The magnetization at low temperature is
larger for the film on STO than it is for the film on LAO. While
this might indicate a change in the total ordered moment, it is not
possible to draw a firm conclusion from such low field data alone.
EXAFS allows for a precise measurement of the Co-O bond length.
Figure 2 displays the R-space Fourier transforms of the Co K-edge
EXAFS spectra for the polarization vector parallel and perpendicular
to the surface normal for the two LSCO films, along with the spectra
for a reference powder of LSCO. The peak in the Fourier transform
near 1.45 \AA{} corresponds to the back scattering from the first
coordinate neighbor O atoms. The peak between 3 and 4 ~\AA{} in Fig.
2 consists of the scattering paths of second shell Co-La and Co-Sr,
and third shell Co-Co along with multiple scattering variants such
as Co-O-Co.

Figures 3 (a) and (b) show the inverse FT of the first-shell
filtered EXAFS for the two LSCO films that correspond to the
first-shell Co-O bond. The difference in the oscillation frequency
between in-plane and out-of-plane is due to the difference of the
Co-O bond lengths. For the film on STO, the in-plane frequency is
smaller than that of out-of-plane, because the Co-O bond length is
longer for in-plane. The situation is reversed for the film on LAO.
Clearly, the Co-O bond lengths adjust to accommodate the strain,
which is consistent with the XAS results of strained Mn-O bond
length in manganite films~\cite{sn04}. In order to quantitatively
determine these bond lengths, the Co-O radial shell for thin films
were modeled with the EXAFS phase and amplitude functions obtained
from the bulk LSCO powder. Assuming the Co-O bond length for powder
is 1.915~\AA{}, we find the bond length of $d_{in}=1.936\pm
0.009$~\AA{} and $d_{out}=1.899\pm 0.013$~\AA{} for the LSCO thin
film on STO, while for the film on LAO, the bond lengths are
$d_{in}$=1.910$\pm$ 0.010~\AA{} and $d_{out}$=1.925$\pm$
0.008~\AA{}, respectively, where $d_{in}$ and $d_{out}$ denotes the
in-plane Co-O bond length and the out-of-plane bond length,
respectively. Alternatively, fitting the out-of-plane data with the
in-plane data for each film, we arrive at the following bond length
shears are $d_{in}-d_{out}$= 0.038$\pm$ 0.009~\AA{} for the LSCO
film on STO substrate and $d_{in}-d_{out}$ = -0.016$\pm$ 0.004~\AA{}
for the film on LAO substrate, respectively. Errors in bond length
were estimated by the spread of distances that resulted in a
doubling of the $\chi^2$ error.

\begin{figure}
\centering
\epsfig{file=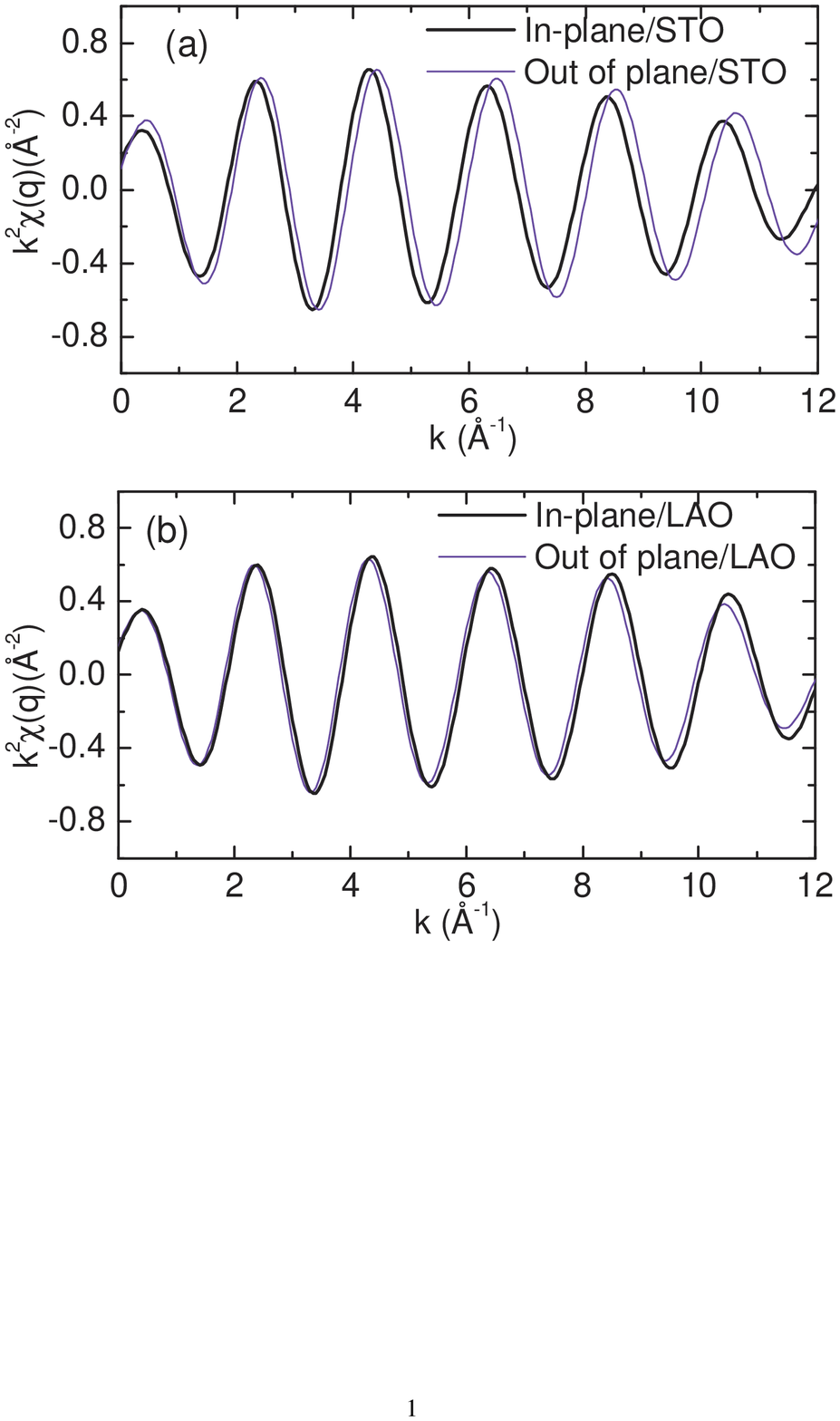,width=6cm,clip=}
\caption{\label{fig:fig3.eps} Inverse FT for the filtered
first-shell Co-O contribution to the Co K-edge EXAFS for LSCO films
grown on (a) STO, and (b) LAO substrates, respectively.}
\end{figure}

\begin{figure}
\centering
\epsfig{file=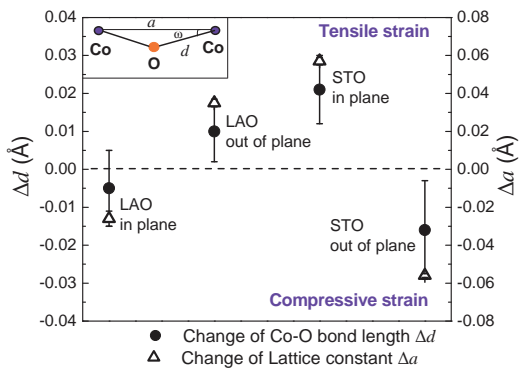,width=7cm,clip=}
\caption{\label{fig:fig4.eps} Changes of Co-O bond length and
lattice constant measured with EXAFS and XRD experiments,
respectively. Error bars are as indicated. The inset shows the
geometrical relation between the bond length $d$, lattice constant
$a$, tilt angle $\omega$, and bond angle $\phi$, which is used to
calculate the relative distortions.}
\end{figure}

An examination of the relative modification of Co-O bond length and
lattice constant allows us to assess the change of bond angle
induced by strain. Figure 4 plots the relative changes of the Co-O
bond length from EXAFS along with the relative changes of the
lattice constant from XRD in both directions for both samples. The
difference in scale for the two vertical axes means that if all of
the strain were taken up in the bond length, the two data points
would coincide. If all of the strain were accommodated by the bond
angle, $\bigtriangleup$$d$ would be fixed at zero. Clearly both are
changing. The inset in Fig. 4 shows the geometrical relation between
the bond length $d$, lattice constant $a$, and tilt angle $\omega$
such that $a = 2 d \cos\omega$, valid for a nearly cubic system.
Assuming small distortions, we derive the change of the tilt angle
in strained thin films,
$\bigtriangleup$$\omega=(2\bigtriangleup$$d$-$\bigtriangleup$$a)/2d\omega$,
and $\bigtriangleup$$\phi=-2\bigtriangleup$$\omega$. Note that bulk
LSCO has the bond angle $\phi$ approximately equal to
168$^{\circ}$~\cite{Caciuffo}. For the film on STO, we calculate
that the in-plane bond angle $\phi$ increases to 172.3$^{\circ}$
$\pm$ 2.6$^{\circ}$, while out-of-plane $\phi$ decreases to
161.2$^{\circ}$ $\pm$ 3.7$^{\circ}$. Similarly, for the film on LAO,
$\phi$ in-plane decreases to 163.5$^{\circ}$ $\pm$ 2.8$^{\circ}$,
while for out-of-plane $\phi$ increases to 172.3$^{\circ}$ $\pm$
2.3$^{\circ}$. Using the bandwidth equation, the full change in the
in-plane bandwidth for the film on LAO compared to the film on STO
is +3.9\%. Consideration of the bond angle alone would give a change
of -0.8\% while consideration of bond length alone would give a
change of +4.7\%. While both factors clearly matter, the bond-length
change is the dominant effect. This is contrary to the assumptions
generally used in the literature on strained oxide films, and in
particular it is different from results reported for manganite
films~\cite{Yuan, am01, hl98}.

The detailed structural information and the measured Curie
Temperatures allow us to examine what models are appropriate for
describing the magnetism in LSCO. The most popular model for
magnetism in these materials is the DE model. In this model, the
exchange energy $J$ is proportional to the electron transfer
integral $t$ or bandwidth $W$. Generally, one finds $T_C \propto J$.
As an extreme but mathematically simple case, we can consider that
$T_C$ is controlled only by the in-plane $J$. Under this assumption
we find that the total percentage change in bandwidth from films on
LAO to those on STO is $\Delta W \simeq$ 3.9\%, considerably smaller
than the change of $T_C$($\simeq$19.2\%). Therefore, the dependence
$J \propto W$ is not large enough to quantitatively account for the
variation observed, with a stronger dependence like $J \propto W^2$
needed. Another mechanism must either supplement or replace the DE
model. For the manganites, an additional possibility discussed by
Millis et al.~\cite{millis} is a static Jahn-Teller (JT) distortion
occurring at the transition~\cite{Radaelli}. That appears not to be
operative for LSCO as there is only a small change in resistivity at
$T_C$, and no sudden change in the JT distortion. A possible
alternative may be found in some early work by
Goodenough~\cite{jb58}. He proposed that the interaction between
Co$^{4+}$ and Co$^{3+}$ could be of a modified superexchange type,
which is regarded as an alternative explanation of ferromagnetism in
La$_{1-x}$Sr$_x$CoO$_3$. Superexchange interactions involve virtual
hopping processes and one expects a larger dependence of the
exchange energy $J$ on the bandwidth $W$, typically $J\propto W^2$,
and thus might better match the observation reported here. However,
superexchange interactions are typically considered to involve
well-localized electrons and thus usually appear for insulators. A
theoretical development explicitly tailored for LSCO will be
necessary to fully understand the microscopic origin of magnetism in
these compounds.



This work is supported through NSF DMR-0239667. Experiments
performed at the National Synchrotron Light Source, Brookhaven
National Laboratory are supported by the U. S. Department of Energy,
Division of Materials Sciences and Division of Chemical Sciences.


\end{document}